\documentclass[12pt,preprint]{aastex}

\slugcomment{To appear in ApJ}

\shorttitle{Dust clearing}
\shortauthors{Brown et al.}

\begin{document}


\title{Evidence for dust clearing through resolved
submillimeter imaging}


\author{{J.M. Brown\altaffilmark{1}, 
G.A. Blake\altaffilmark{2}, C. Qi\altaffilmark{3}, C.P. Dullemond\altaffilmark{4}, D.J. Wilner\altaffilmark{3}, J.P. Williams\altaffilmark{5}}}
\altaffiltext{1}{Max-Planck-Institut f{\"u}r extraterrestrische Physik, Garching bei M{\"u}nchen, Germany; jbrown@mpe.mpg.de}
\altaffiltext{2}{Division of Geological \& Planetary Sciences, MS150-21, California Institute of Technology, Pasadena, CA 91125}
\altaffiltext{3}{Harvard-Smithsonian Center for Astrophysics, 60 Garden Street, Mail Stop 42, Cambridge, MA 02138}
\altaffiltext{4}{Max-Planck-Institut fur Astronomie, K{\"o}nigstuhl 17,
69117 Heidelberg, Germany}
\altaffiltext{5}{Institute for Astronomy, 2680 Woodlawn Drive, Honolulu, HI 96822 }



\begin{abstract} Mid-infrared spectrophotometric observations have
  revealed a small sub-class of circumstellar disks with spectral
  energy distributions (SEDs) suggestive of large inner gaps with low
  dust content. However, such data provide only an indirect and
  model-dependent method of finding central holes. Imaging of
  protoplanetry disks provides an independent check of SED
  modeling. We present here the direct characterization of three 33-47
  AU radii inner gaps, in the disks around LkH$\alpha$ 330, SR 21N and
  HD 135344B, via 340 GHz (880 $\mu$m) dust continuum aperture
  synthesis observations obtained with the Submillimeter Array
  (SMA). The large gaps are fully resolved at $\sim$0\farcs3 by the
  SMA observations and mostly empty of dust, with less than 1 - 7.5
  $\times$ 10$^{-6}$ M$_\odot$ of fine grained solids inside the
  holes. Gas (as traced by atomic accretion markers and CO 4.7
  \micron\, rovibrational emission) is still present in the inner
  regions of all three disks. For each, the inner hole
  exhibits a relatively steep rise in dust emission to the outer disk,
  a feature more likely to originate from the gravitational influence
  of a companion body than from a process expected to show a more
  shallow gradient like grain growth.  Importantly, the good agreement
  of the spatially resolved data and spectrophotometry-based models
  lends confidence to current interpretations of SEDs, wherein the
  significant dust emission deficits arise from disks with inner gaps
  or holes. Further SED-based searches can therefore be expected to
  yield numerous additional candidates that can be examined at high
  spatial resolution.

\end{abstract}




\section{Introduction}

Knowledge of how disks dissipate is vital for our understanding of
planetary system formation. Studies have shown that warm dust in
the inner regions of massive, optically thick disks survives for a 
period of order $\sim$3-5 Myr (e.g. \citealt{strom89}; 
\citealt{beckwith90}). Very few objects in transition between this
primordial state and the optically thin debris disk systems first
discovered by IRAS (\citealt{aumann84}; \citealt{rieke05})
are known, making difficult the identification of a sample of this
important, short-lived phase. One indicator of intermediate
systems is the presence of an inner hole or gap indicating that the
inner disk has evolved while the outer disk has not. The presence of
gaps may indicate that planets have already formed in the disks and
cleared the material around their orbits. Such systems can therefore
strongly constrain models of planet formation, especially the role of
gap formation and disk-planet interactions in various planet migration
scenarios that lead to the creation of ``hot Jupiters'' found to orbit
much older systems \citep{marcy05}. It is therefore essential to
search for other types of evidence that either support or reject the
gap hypothesis.

In theory, an inner gap in a proto-planetary disk can be identified
from a depressed SED at wavelengths of 1-15 \micron, as the absence of
hot dust close to the star results in flux coming solely from the
stellar photosphere, rather than disk surface emission. To date, such
emission ``deficits'' are the tool most widely used to infer the
presence of gaps (\citealt{brown07}, \citealt{calvet02},
\citealt{forrest04}), but spectrophotometric signatures are indirect
and notoriously difficult to interpret as multiple physical scenarios
can result in the same SED \citep{boss96}. Additional constraints and
detailed models are therefore required to distinguish between the
possible physical scenarios that are consistent with the observed
fluxes.

The relatively new field of submillimeter interferometry provides
access to the small scales necessary for examining inner disk
structure. Submillimeter imaging is optimal for transitional disk
studies as the sources are intrinsically stronger than at longer
centimeter and near-millimeter wavelengths. The spatial filtering
possible with arrays can also suppress the contamination produced by
the weak extended emission from nearby molecular cloud material that
is routinely picked up by single dish measurements of the
submillimeter flux. Long baselines and good $(u,v)$ coverage are vital
to provide the high resolution/high dynamic range imaging needed to
resolve the gaps. Because submillimeter observations trace optically
thin emission in the Rayleigh-Jeans regime, the data are exquisitely
sensitive to the mass surface density profile in the disk. However,
only the brightest disks are suitable for resolving disk structure and
few transitional disks have been studied on the scales necessary to
resolve the central hole (LkCa 15 - \citealt{pietu06}; TW Hya -
\citealt{hughes07}; LkH$\alpha$ 330 - \citealt{brown08}; GM Aur -
\citealt{dutrey08}, \citealt{hughes09}; SR 21N - \citealt{andrews09}).

Here we present further examples supporting the gap
hypothesis in the form of 340 GHz (880 \micron) continuum maps
resolving the inner disk holes in the disks around LkH$\alpha$ 330,
SR21 and HD 135344B. Details about the sources and relevant observational
details are presented in Section 2. We then turn to a discussion of the 
results and their implications for SED-driven searches for gaps in 
circumstellar disks.

\section{Observations}

Mid-IR spectrophotometry of LkH$\alpha$ 330, SR 21N and HD 135344B was
acquired as part of the Spitzer ``From Cores to Disks'' (c2d) Legacy
Science project. Out of a sample of over 100 spectra in the c2d first
look program, chosen from known T Tauri stars, only 5 disks
(LkH$\alpha$ 330, SR 21 N, HD 135344 B, T Cha, and CoKu Tau 4) showed
SED features characteristic of an inner hole \citep{brown07}. Those
disks visible from the northern hemisphere and with suitably large
disk masses -- LkH$\alpha$ 330, SR 21N and HD 135344B -- have been targeted
for high spatial resolution follow-up imaging at submillimeter
wavelengths.

LkH$\alpha$ 330 is a little studied G3 star near the IC 348 region of 
Perseus, which is a sparse cluster with stars between a few to ten 
million years old \citep{ssc74}. The distance
to Perseus is an unresolved problem with values ranging from 200 to
350 pc. A distance of 250 pc is assumed here following the c2d convention.

SR 21N (a.k.a. Elias 2-30) is a 3 Myr old pre-main sequence binary,
with a separation of 6\farcs4, in the core of the $\rho$ Ophiucus
cloud at a distance of 160 pc \citep{pgs03}. The primary has a
spectral type of G2.5, while the companion has spectral type M4. A
recent VLT/NACO AO survey found no other companions
\citep{correia06}. Interestingly, \citet{pgs03} found that the two
companions were not coeval within their limits, although large
uncertainties remain. The disks of the two components are closely
aligned indicating that the stars are likely gravitationally bound but
the distance between the two is large enough that we do not expect
significant gravitational perturbations to the SR 21N disk,
particularly in the inner regions.  \citep{jensen04}.

HD 135344B is an 8 Myr old F4 star in Lupus that lies $\sim$20
arcseconds from its A-type companion HD 135344A. The two stars are
likely not physically associated. The dust disk around HD 135344B has
been spatially resolved in UV scattered light \citep{grady05,grady09} and the
mid-IR \citep{doucet06}. However, the two observations provide
different disk inclinations of $<$20$^\circ$ \citep{grady09} and 46$^\circ$
\citep{doucet06}. A close (0\farcs32 separation) binary system lies
5\farcs8 to the southwest \citep{augereau01}. A distance of 84 pc in
agreement with \citet{dunkin97} was used in \citet{brown07}. A further
distance of 140 pc has been suggested by \citet{vanboekel05} and
provides a better simultaneous fit of the SMA image and SED.

\begin{figure} \begin{minipage}{0.9\linewidth} 
\begin{center}
\includegraphics[angle=-90,scale=0.3]{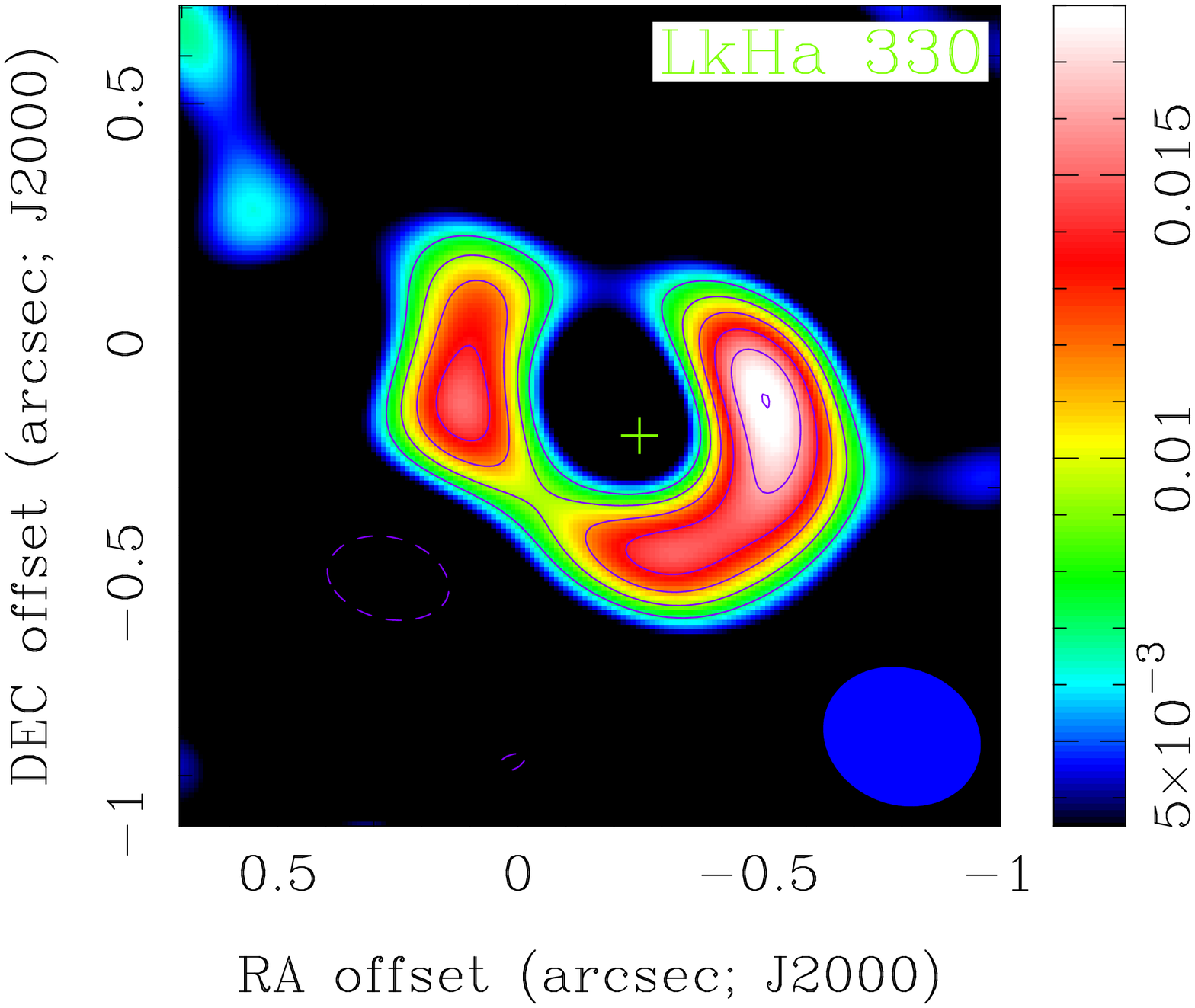}\\
\includegraphics[angle=-90,scale=0.3]{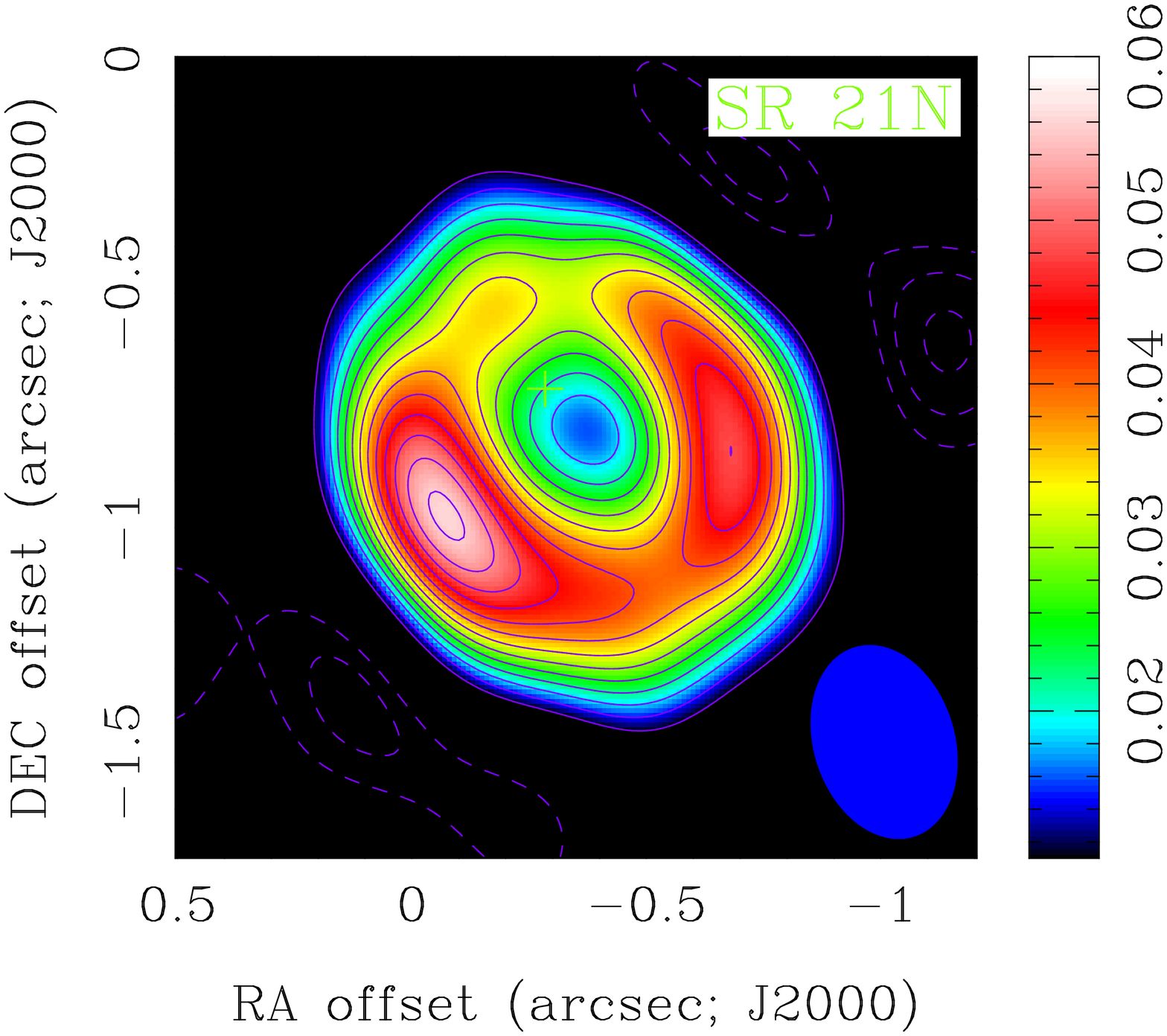} \\
\includegraphics[angle=-90,scale=0.3]{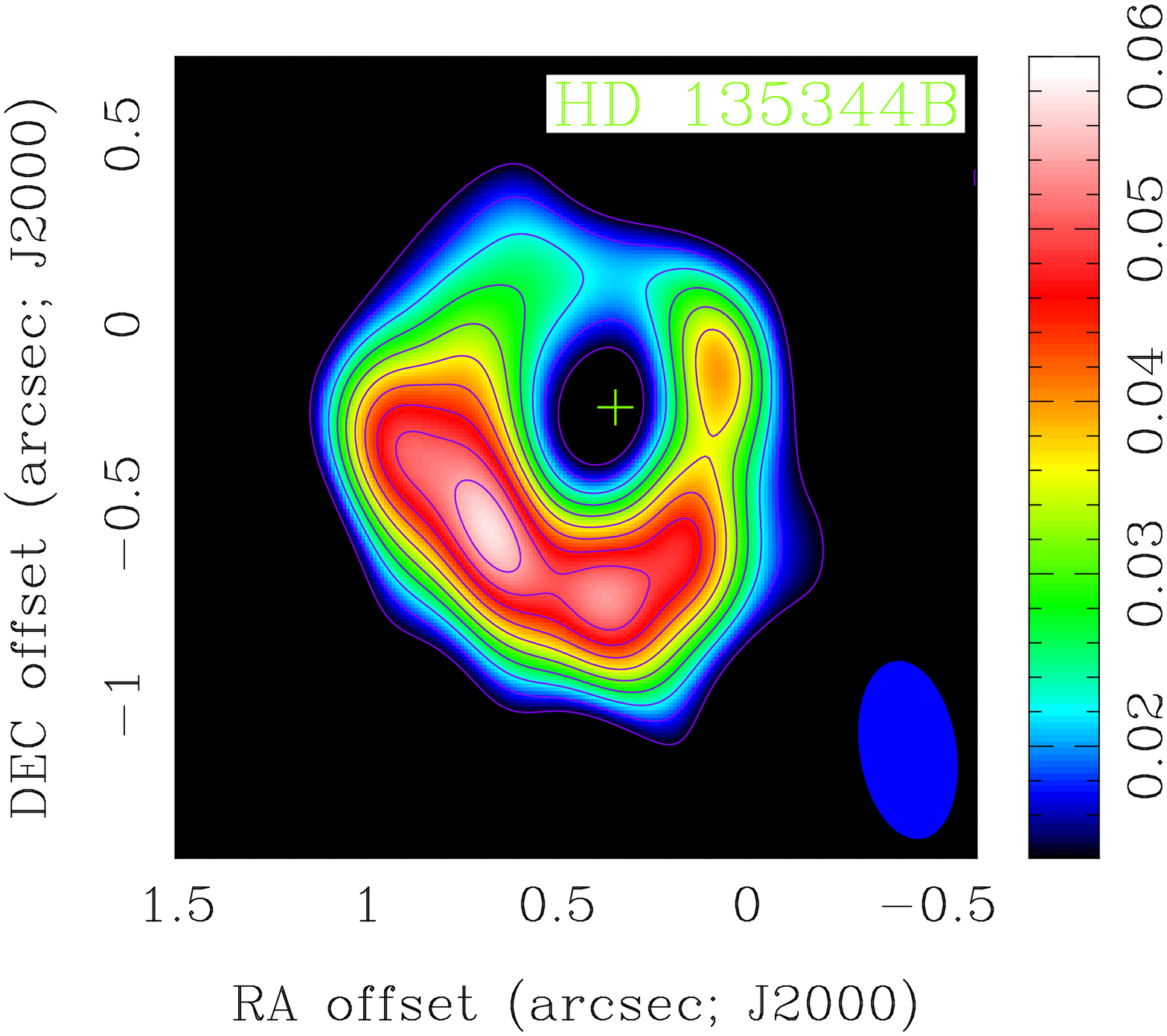}
\end{center}
\caption{The 340 GHz dust continuum images of LkH$\alpha$ 330 (top),
SR 21N(middle) and HD 135344B (bottom). The crosses mark the
literature coordinates of the central star. LkH$\alpha$ 330 clearly
shows an inner hole of approximately 40 AU radius with the synthesized
beam of 0\farcs28x0\farcs33 (plotted at bottom right).  SR 21N has the
smallest hole of this sample with a radius of 27
AU. HD 135344B is
the most asymmetric of the disks and has a well defined 37 AU
hole. The 0\farcs47x0\farcs25 beam (lower right corner) is elongated
due to HD 135344B's -37$^\circ$ degree declination.\label{fig:lkha330sma}}
\end{minipage} 
\end{figure}

Dust emission measurements were acquired with the Submillimeter Array
(SMA) using the very extended configuration with the eight 6 meter
diameter antennae, which provided baselines ranging in length from 30
to 590 m. The observations of LkH$\alpha$ 330 in 2006 November were
taken with only seven antennae and the minimum length baseline was 80
m. Double sideband (DSB) receivers tuned to 341.165 GHz provided 2 GHz
of bandwith/sideband, centered at an Intermediate Frequency (IF) of 5
GHz. Calibration of the visibility phases was achieved with
observations of a quasar within 10 degrees of the source (listed in
Table \ref{table:smaobs}), typically at intervals of 25 minutes. 3C273
was used as the passband callibrator in all observations. Measurements
of one of Uranus, Titan and Callisto provided the absolute scale for
the flux density calibration.  Uncertainties in the flux scale are
estimated to be 15\%. Table \ref{table:smaobs} lists the observed
source position, the synthesized beam size with natural weighting,
single DSB system temperatures, the phase and flux callibrators and
the observation dates. Comparison of the millimeter source positions
with the optical stellar positions (marked with crosses in Figure
\ref{fig:lkha330sma}) show that within the position errors the
millimeter emission is centered on the star. SR 21 N has the least
accurate optical position and has the largest difference between the two
positions.  HD 135344B and SR 21N have higher system temperatures and
more elongated beams than LkH$\alpha$ 330. The data were calibrated
using the MIR software package ({\tt
  http://cfa-www.harvard.edu/$\sim$cqi/mircook.html}), and processed
with Miriad \citep{sault95}.

\section{Data}

\subsection{Image plane}

The SMA images clearly resolve the size, orientation and radial
structure in all three disks (see Figure \ref{fig:lkha330sma}). The
parameters found directly from the data are summerized in Table
\ref{table:smaimages}. The total fluxes, $F_{\rm total}$, were found
by integrating all disk flux above a 3-sigma noise cutoff and the peak
S/N ranged from 25-75 for the three disks (Table
\ref{table:smaimages}, columns 2 and 3). The hole radii, $R_{\rm
  hole}$, (column 4) are in good agreement with the SED-determined
values. The outer disk edge, $R_{\rm disk}$, is also resolved. Due to
the lower optical depths in the outer disk, this high resolution
imaging is likely not sensitive to the outermost reaches of the disk,
so the radii in column 6 are lower limits. For all three disks, upper
limits on the integrated flux within the hole, $F_{\rm hole}$, were
derived (column 7). The SR 21N hole is smaller than those in the other
disks and barely resolved, so the flux decrement is not as well
constrained.  In the LkH$\alpha$ 330 image, the flux within the hole
is below the 1 sigma noise level of 2.3 mJy/beam. Using the flux to
disk mass conversion of \citet{beckwith90}, which implicitly assumes a
gas to dust ratio of 100, limits on the amount of mass in the hole on
the order of 10$^{-4}$ M$_\odot$ were calculated (column 8).  In the
case of LkH$\alpha$ 330, the significant intensity contrast between
the inner and outer disk indicates a large mass surface density
contrast -- even for mm-sized grains.

The millimeter inclinations were found by assuming a circular disk and
deprojecting the minor axis after correcting for the beam. The unusual
geometries and asymmetries introduce fairly large systematic errors
into these results. In the case of HD 135344B, the inclination agrees
better with HST scatter light images of the outer disk \citep{grady09}
than resolved 20\micron images \citep{doucet06}. The large asymmetry,
seen also in the HST images, is a potential cause of the inclination
discrepancy, particularly for inclinations derived from lower spatial
resolution data.

\subsection{(u,v) plane}

\begin{figure}
\begin{center}
\includegraphics[angle=0, scale=0.5]{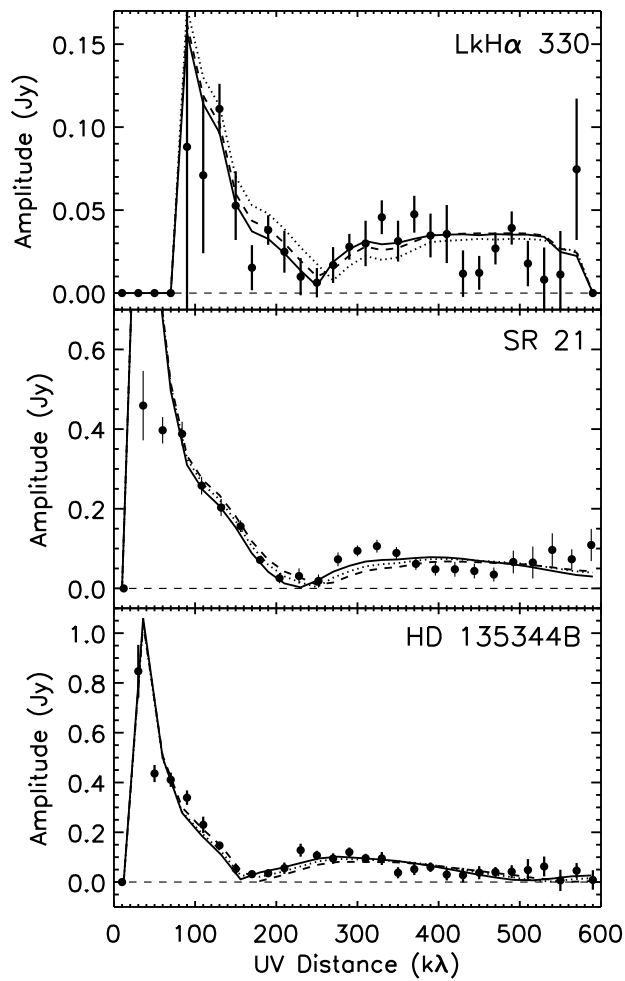}
\end{center}

\caption{Amplitude vs (u,v) distance for, from top to bottom,
  LkH$\alpha$ 330, SR 21N, and HD 135344B. The data are marked with
  filled circles and the step-function model with a solid line. One
  sigma error bars are included on the data points. The dotted and
  dashed fits show models with gap transitions $R_{\rm Edge}/R{_{\rm
      Hole}}$ of 0.25 and 0.5 respectively. The horizontal dashed line
  marks zero amplitude. The (u,v) coverage for LkH$\alpha$ 330 is
  sparser due to the missing antenna resulting in larger error bars
  and some poorly sampled bins. \label{fig:smauv}}

\end{figure}

With a sufficiently sharp transition, aperture synthesis observations
should detect a null in the flux versus ($u,v$)-distance for disks
with gaps, as opposed to the smooth drop off in flux associated with
power-law mass surface density profiles that characterizes most
classical T Tauri star disks (\citealt{aw07}). A visibility domain
analysis also provides a more straightforward means of assessing the
uncertainties in any fits, as the data have not been affected by the
Fourier transforms and non-linear deconvolution inherent to CLEAN/MEM
routines. The visibility data were averaged in concentric ellipses of
deprojected (u,v) distance to account for disk inclination, $i$, and
position angle, PA (e.g. \citealt{hughes07}). Thus, the deprojected (u,v) distance is $R =
\sqrt{d_{\rm a}^2 + d_{\rm b}^2}$ with $d_{\rm a}= {\rm R \, sin}
\phi$ and and $d_{\rm b}= {\rm R\, cos} \phi \, {\rm cos}\, i$ where
$\phi={\rm arctan(v/u - PA)}$. The disks are all close to face-on and
deprojecting the (u,v) distance does not make a large difference
compared to a simpler annular average. All three disks do indeed show
nulls (see Figure \ref{fig:smauv}), the values of which are listed in
Table \ref{table:smaimages}. 

\subsection{Spectral energy distributions}

\begin{figure}
\begin{minipage}{0.9\linewidth}
\hspace{-1cm}
\includegraphics[angle=0, scale=0.5]{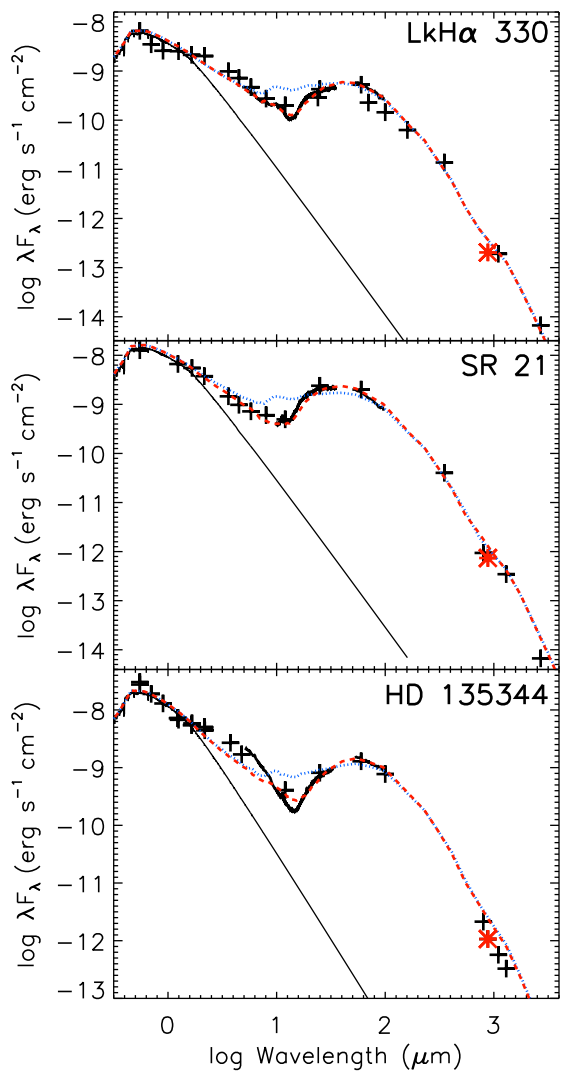}
\end{minipage}
\caption{Model fits to the SEDs of LkH$\alpha$ 330 (top), SR 21N
  (middle) and HD 135344B (bottom). The models used to fit the images
  (dashed red line) are overlaid on the photometry (crosses) and IRS
  spectra (black line), confirming that the models fit the SEDs as
  well as the images. The dotted blue line is the equivalent model
  with no hole and the solid black curve is the stellar
  photosphere. The SMA total fluxes have been placed in the SED as a
  red star, and are consistent with previous photometry. \label{fig:smaseds}}
\end{figure}

The dust grains that emit at 10 $\mu$m versus 1 mm are very different
in size, and may therefore arise from very different populations.
Thus, it is of interest to examine in detail whether the SED-derived
disk structure is in agreement with that dictated by the SMA images.
The total 880 \micron\, flux from the SMA observations was included in
the photometry to check the overall flux in relation to other
measurements (see the star-labeled points in Figure
\ref{fig:smaseds}). The integrated fluxes do fit well with the other
submillimeter measuements in the log-log plots. On average, however,
the fluxes are slightly lower than other data, and the (u,v)-plane
data indicate a flux deficit on the shortest spacings as compared with
the model. This is likely due to the lack of short baselines in the
very extended SMA configuration. Fits to the optical through
millimeter-wave SED yield estimated gap outer radii for all three
disks between 20 and 50 AU (see \citealt{brown07}), which are roughly
in agreement with the images. The largest discrepancy is in SR 21
where the millimeter hole is significantly bigger than that found from
fitting the SED. Such differences could arise either from incorrect
parameterization of the central star in the model or from an intrinsic
difference in the distribution of different sized grains.

\section{Modeling}

\begin{figure*}
\begin{minipage}{0.5\linewidth}
\begin{center}
\includegraphics[angle=-90, scale=0.3]{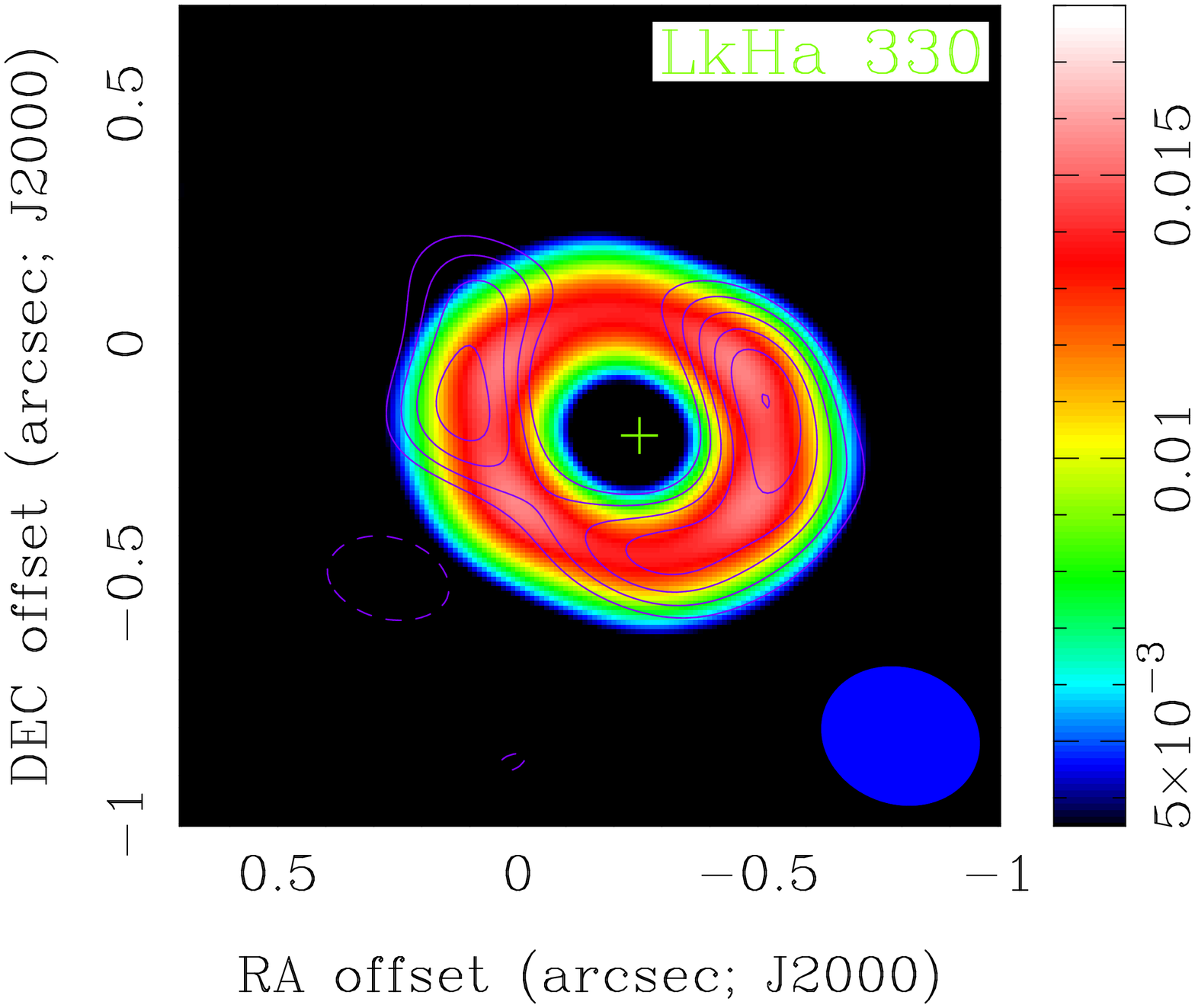}
\end{center}
\end{minipage}
\begin{minipage}{0.5\linewidth}
\begin{center}
\includegraphics[angle=-90, scale=0.3]{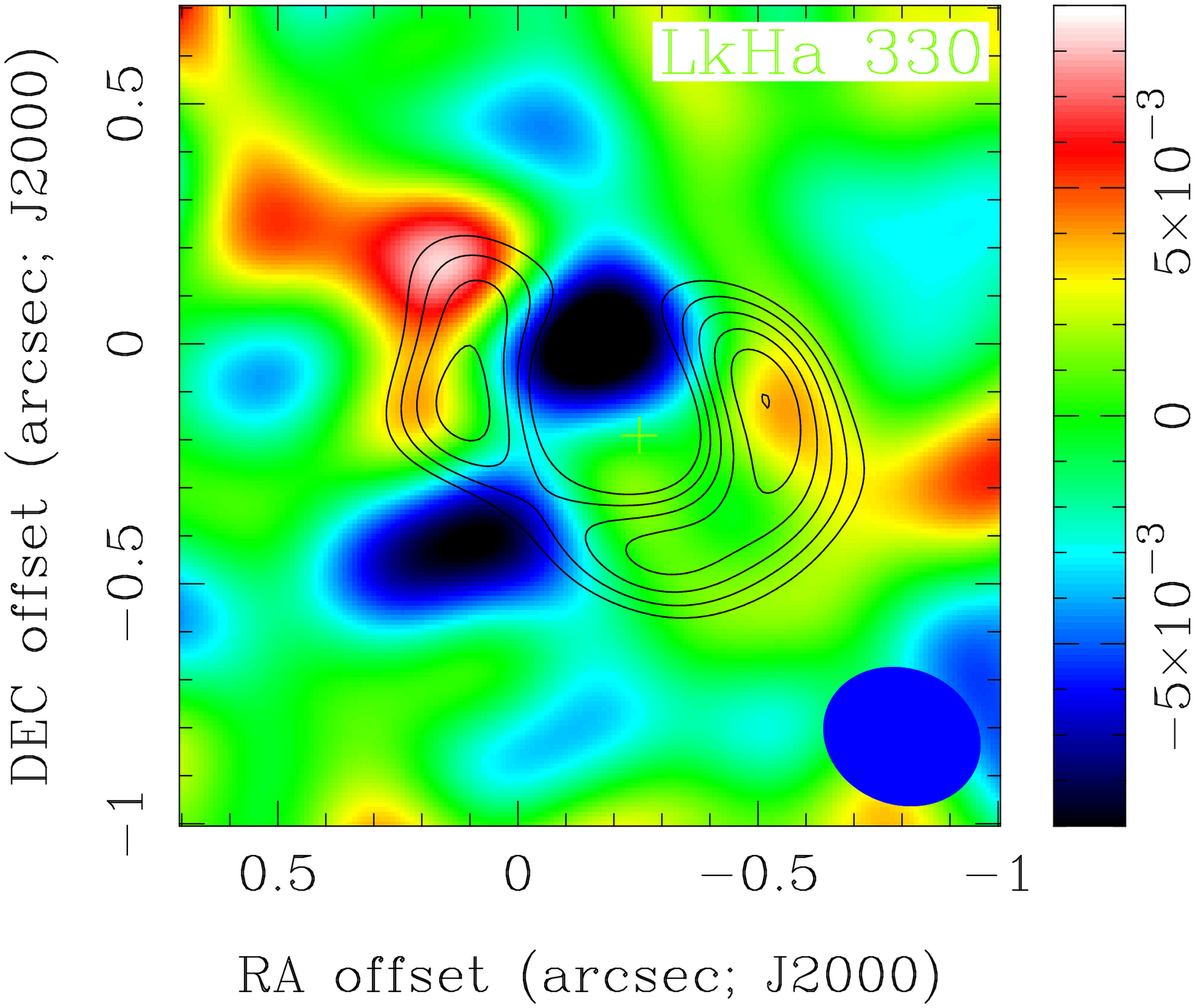}
\end{center}
\end{minipage}

\begin{minipage}{0.5\linewidth}
\begin{center}
\includegraphics[angle=-90, scale=0.3]{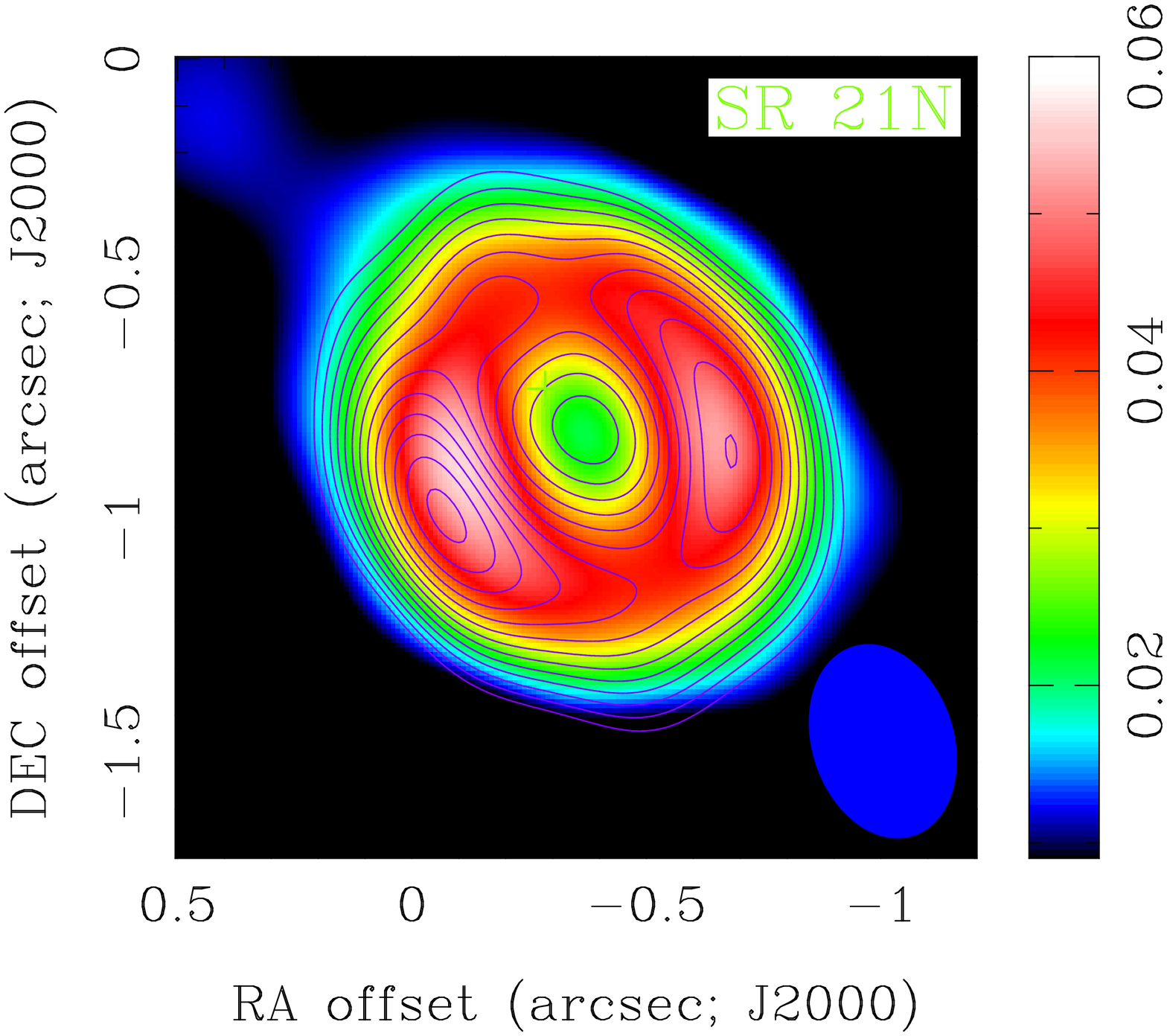}
\end{center}
\end{minipage}
\begin{minipage}{0.5\linewidth}
\begin{center}
\includegraphics[angle=-90, scale=0.3]{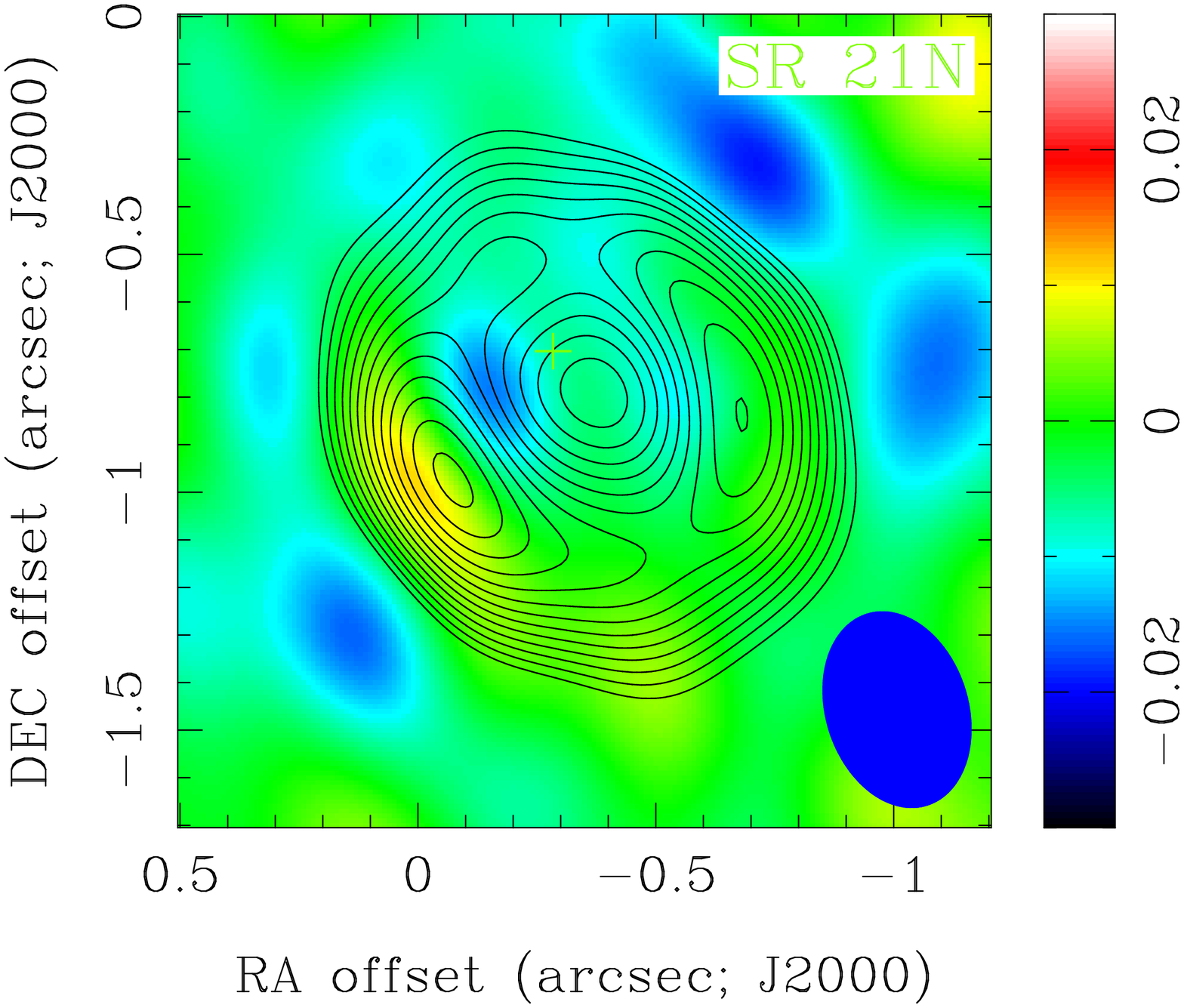}
\end{center}
\end{minipage}

\begin{minipage}{0.5\linewidth}
\begin{center}
\includegraphics[angle=-90, scale=0.3]{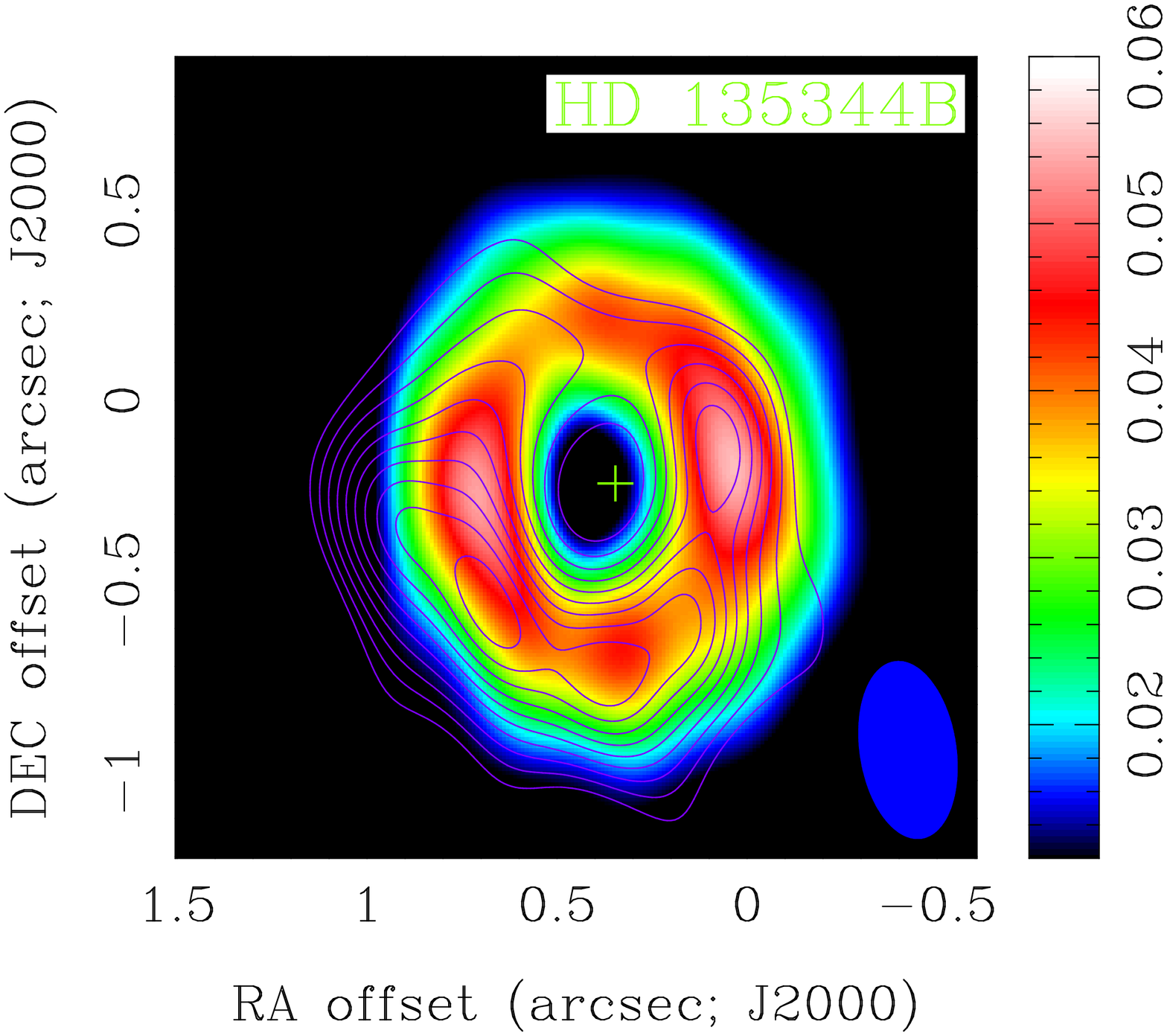}
\end{center}
\end{minipage}
\begin{minipage}{0.5\linewidth}
\begin{center}
\includegraphics[angle=-90, scale=0.3]{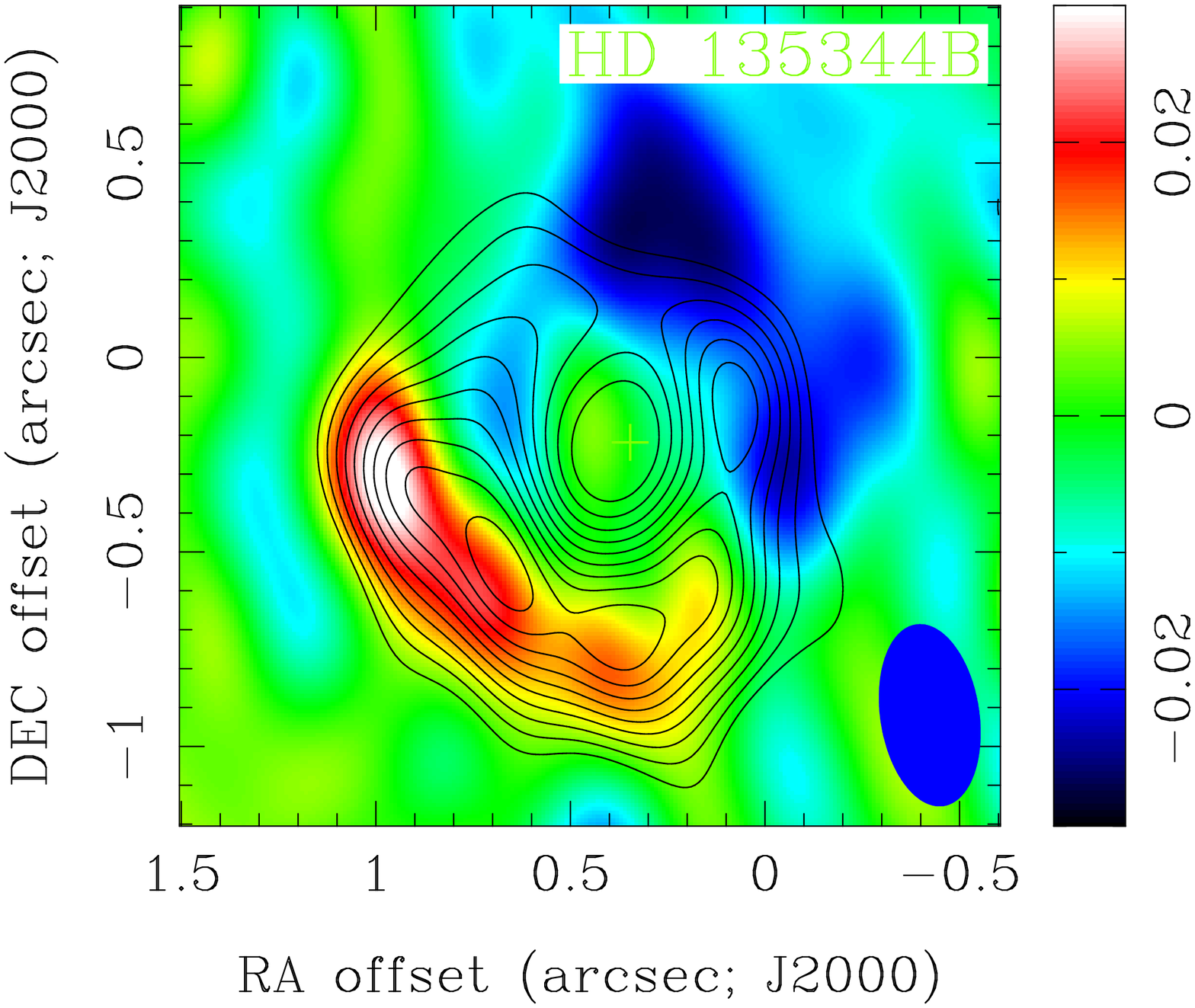}
\end{center}
\end{minipage}
\caption{On the left are the models of LkH$\alpha$ 330 (top), SR 21N
  (middle) and HD 135344B (bottom) disk 880 $\mu$m emission in color
  overlaid with 1-sigma contours from the data, beginning at 3-sigma.
  The synthesized beams are depicted in the lower right corners and
  the stellar positions are marked with a cross. The model determines
  the hole radii to be 47 AU for LkH$\alpha$ 330, 33 AU for SR 21N and
  39 AU for HD 135344B. On the right are the residuals when the model
  is subtracted from the data. Dark regions are areas where the model
  overproduces flux and light regions are areas where the model
  underproduces flux. The scales are set such that the extremes of the
  residual contours are 50\% of the peak flux. \label{fig:smamodels}}
\end{figure*}

The 2-D radiative transfer model RADMC \citep{dd04}, as modified
to include a density reduction simulating a gap, was used to simultaneously
model the resolved images (see Figure \ref{fig:smamodels}), the data
in the (u,v)-plane (see Figure \ref{fig:smauv}), and the SEDs (see
Figure \ref{fig:smaseds}). This model assumes a passive disk, which
merely reprocesses the stellar radiation field. In order to fit the
missing dust emission, the model was adapted to reduce the dust
density in a specific region to create a gap in the disk. The
resulting image was resampled in Miriad using the same (u,v)-plane
distribution as the SMA data so the two are directly comparable (see
Figure \ref{fig:smamodels}).

In the model, the disk is assumed to be flared such that the surface
height, $H$, varies with radius, R, as $H/R \propto R^{2/7}$, as in
\citet{cg97}. As in \citet{brown07}, the models do not calculate the
hydrostatic equilibrium self-consistently, and the pressure scale
height is anchored at the outer disk edge, in this case 300 AU, in
agreement with values predicted from hydrostatic equilibrium.  Density
is a power law of radius with an index of -1. The dust composition is
set to a silicate:carbon ratio of 4:1 with only amorphous, rather than
crystalline, silicate included. The grain sizes range from 0.01 $\mu$m
to 10 cm with a power-law index of -3.5 and a total disk mass,
including both gas and dust, of $\sim$0.02 M$_\odot$ \citep{ob95}. The
inner edge of the disk, $R_{\rm Disk,in}$, was set at approximately
the radius where the dust sublimation temperature, 1500 K, is reached.

The gap is represented in the model by three parameters: both an inner
and an outer gap radius and a density reduction factor. The models
were determined based primarily on the SED models found using $\chi^2$
minimization on the inner and outer gap radii in
\citet{brown07}. However, in the case of HD 135344B the SED model at
84 pc resulted in a hole of 1\farcs1, significantly larger than that
seen in the SMA image. In order to resolve this discrepancy, a revised
distance of 140 pc and increased stellar luminosity was used (as in
\citealt{vanboekel05}), providing good agreement between the SED and
image. The SED best fit models matched the images remarkably well,
lending confidence to current SED interpretation and modeling. Both
the image and the SED indicate that the gap in the LkH$\alpha$ 330
disk has an outer radius of 47 AU. HD 135344B has a hole radius of 39
AU, while SR 21N has the smallest hole with a radius of 33 AU. The
null in the (u,v) plane provides the strongest constraint on the
model, limiting good fits to within $\pm$4 AU or a tenth of a
beam. Large differences in the initial model parameters or distances
could result in larger errors. The SMA images places no constraint on
the inner gap radius, but the SEDs require hot dust with temperatures normally only found at $<$1 AU. The models thus have a narrow ring of hot gas extending to 0.45 AU for SR 21N and HD 135344B and 0.8 AU for LkH$\alpha$ 330. Within our solar system, this gap corresponds to
the area between Earth's orbit and distances just beyond Pluto's
orbit.

\subsection{Density contrast}

The LkH$\alpha$ 330 hole is largely empty of dust, and the intensity
contrast between the hole and the outer disk is large, indicating a
similarly large mass surface density contrast. The best fit models for
all three disks have density reductions of 1000 within the hole. In
the LkH$\alpha$ 330 disk, this corresponds to 2.1 $\times$ 10$^{-5}$
M$_\odot$ of material, compared to less than 1.3 $\times$ 10$^{-4}$
M$_\odot$ derived from the image. For comparison, the model without
the hole contains some 0.01 M$_\odot$ of material within the same
region. SR 21N and HD 135344B, with their more gradual gap edges, have
less well defined density contrasts and more material within the
holes. Limits on the amount of material in the hole are 3.2 $\times$
10$^{-5}$ M$_\odot$ and 2.9 $\times$ 10$^{-5}$ M$_\odot$,
respectively. 

\subsection{Gap edge}

For all of the disks, the inner hole exhibits a relatively steep rise
in dust emission to the outer disk and can be modelled satisfactorily
with a step function density reduction. In order to investigate the
abruptness of the transition between the inner and outer disk, a gap
edge was introduced so that the density reduction rises
logarithmically over a range of radii, $R_{\rm Edge}$, around the
outer gap radius. This effect is similar to moving the hole radius
inwards but over large scales the two disk structures can be
distinguished. Figure \ref{fig:smauv} shows models with different
values of $R_{\rm Edge}/R_{\rm Hole}$ compared to the
data. LkH$\alpha$ 330 is best fit by the steepest profile as might be
expected from the large intensity contrast between the hole and outer
disk. However, both HD 135344B and SR 21N can be fit well with much
shallower slopes in density reduction, $R_{\rm Edge}/R_{\rm Hole}$ up
to 0.5. A steep transition is more likely to originate from the
gravitational influence of a companion body than from a process
expected to show a more shallow gradient like grain growth.

\subsection{Asymmetries}

The most prominent asymmetry in the images is due to the inclination
of the disks. The edges where more dust lies along the line-of-sight
are brighter, creating two bright regions along the major
axis. However, other significant asymmetries remain which cannot be
explained by an axisymmetric disk model (see Figure
\ref{fig:smamodels}). Such asymmetries could arise from
gravitational perturbation caused by a large planet or binary
companion. 

The disk around HD 135344B is the most asymmetric with an almost
horseshoe shape, so much so that modeling with an axisymmetric disk
can never provide a good fit. The vast majority of the disk material
lies to the south/southeast, with a cleared region in the ring to the
north (see Figure \ref{fig:smamodels}). In order to quantitatively
measure the amount of asymmetry, the ratios of integrated flux on the
two sides of both the major and minor axes were measured. In all cases
the asymmetries are larger around the minor axes. For HD 135344B, the
flux ratio around the major axis is 1.3 but the large visible
asymmetry around the minor axis is 2.2 -- although the difference from
the brightest point to its counterpart on the other side is a factor
of 10. Both SR 21N and LkH$\alpha$ 330 are more symmetric. SR 21N has
a flux asymmetry of 1.1 around the major axis and 1.4 around the
minor. LkH$\alpha$ 330 has only a slight asymmetry around the major
axis and 1.3 around the minor.

\section{Discussion}

We have directly imaged large inner gaps in three young protoplanetary
disks. The observed sizes are in excellent agreement with those found
through SED-based fits. Synthetic images, produced by the 2D radiative
transfer code used to model the SEDs, fit the data well, although the
asymmetry of the HD 135344B disk poses problems with the axisymmetric
model used. The limits from the SMA images on the amount of material
within the holes are also in agreement with the models and reveal a
significant reduction in dust in the inner regions. For all of the disks, the
inner hole exhibits a relatively steep rise in dust emission to the
outer disk.

While the SMA images provide a large amount of information about the
outer regions of the disk, information regarding the structure close
to the star must come from other sources.  The presence of material
close to the star, making the density reduction a gap rather than a
hole, places constraints on many gap producing processes. At present,
the only indications of dust close to the star are the 1-10 \micron\,
excesses. All the models require a hot dust component to fit the
near-IR excess in the SEDs, modeled in this paper as a ring of matter
close to the star. However, the SEDs place no spatial constraints on
the source of this emission. \citet{fedele08} find inner dust between
0.05 AU and 1.8 AU in HD 135344B which is consistent with a ring model
as presented here. On the other hand, \citet{eisner09} find that
mid-IR aperture masking visibilities of SR 21N are incompatible with a
compact ring of emission. They preferentially model this near-IR
excess as a low mass companion. However, there is little positive
evidence of such a companion, with the expected degree of asymmetry
not seen in either their own data or in 5 \micron\, CO data
(\citealt{pontoppidan08}; Pontoppidan, priv. comm.). In either case,
the amount of mass in these scenarios is negligible compared to the
outer disk and cannot be distinguished from the millimeter data
presented here.

There are several indications that gas must reside close to the
stars. First, LkH$\alpha$ 330 and HD 135344B display H $\alpha$
emission, with equivalent width measurements ranging from 11 to 20
\AA\, (\citealt{fernandez95}, \citealt{ck79}), and so are still
accreting gas. Additionally, \citet{vanderPlas08} found [O I] emission
from within the dust gap in HD 135344B. SR 21N, on the other hand,
does not appear to still be accreting or only very weakly
\citep{natta06}. Second, emission from warm (800 - 1000 K) gas is seen
in the 4.7 \micron\, $v$=1$\rightarrow$0 rovibrational emission lines
of CO from all three sources (\citealt{pontoppidan08},
\citealt{salyk09}). \citet{pontoppidan08} spatially resolved the warm
CO emission in SR 21N and HD 135344B using spectro-astrometery, and
placed the inner edge of the gas emission regions at 7 AU and 0.5 AU,
respectively.

Formation of these dust holes by planetary companions would be the
most exciting explanation for their presence, and extensive
theoretical work has been done to fit transitional disk observations
with models of disks with planets (\citealt{varniere06}). Theories
reconciling the presence of gas within the planet-induced holes
include dust filtration \citep{rice06} and gas transport by
magneto-rotational instability \citep{chiang07}. However, other
processes which might form a hole may be effective and must be
considered. Stellar mass companions in close binary orbits remain one
of the most likely causes of inner disk holes and the SEDs show the
defining characteristics of a transitional SEDs (e.g. CoKu Tau 4,
\citealt{forrest04}, \citealt{ik08}; CS Cha,
\citealt{espaillat07b}). Although no companions are currently known in
these disks, low mass binary companions are difficult to detect
without dedicated searches. Interferometric aperture-masking observations
with NIRC2 on Keck reveal that LkHa330 and SR 21 have no companions
with masses of $>$ $\sim$50 M$_{\rm Jupiter}$, at separations larger than
$\sim$10 and $\sim$5 AU, respectively (M. Ireland and A. Kraus,
private communication).

One proposed process for quickly clearing the inner disk region is
photoevaporation (\citealt{clarke01}, \citealt{alexander06}). An inner
hole occurs when the photoevaporation rate driven by the ionizing flux
from the central star matches the viscous accretion rate. However,
this condition is only effectively fulfilled when accretion rates are
low and would result in no gas or dust close to the star for gap radii
of several tens of AU.  Photoevaporation is thus extremely unlikely to be
responsible for the 27-40 AU radii gaps observed in the LkH$\alpha$
330, HD 135344B and SR 21N disks (\citealt{salyk09}).

An alternative explanation to physical removal of the dust is that it
has grown beyond the size at which it efficiently radiates as a
blackbody so that it no longer emits strongly in the mid-IR and
submillimeter \citep{tanaka05}.  Within any realistic distribution of
dust grain sizes, even a minimum grain diameter of 50 \micron\,
significantly overproduces the flux in the 10 \micron\, region with no
density reduction. Thus, grain growth to very large sizes with little
fragmentation in collisions is needed for this scenario to be relevant
for these transitional disks and seems unlikely from a theoretical
point of view \citep{brauer08}. The sharp cutoffs in dust mass surface
density between the inner and outer disk in LkH$\alpha$ 330 and Hd
135344B are also difficult to reconcile with dust coagulation models
alone.

\section{Conclusions}

To summarize, LkH$\alpha$ 330, HD 135344B and SR 21N present dramatic
cases of disks evolving from the inside out rather than smoothly
throughout all radii as would be expected in (constant) alpha-viscosity 
models of disk evolution. The gaps are largely empty of dust but gas does
remain, and the outer edge of the gaps rise steeply. Importantly, the
good agreement of both data and model lends confidence to current
interpretations of SEDs with significant dust emission deficits being
due to inner holes in disks. Further SED-based searches can therefore
be expected to yield numerous additional candidates that can be
examined at high spatial resolution. Ultimately, in such studies it
will be critical not only to image the dust but to provide estimates
of the gas:dust ratios in the outer and inner disk if the different
possible gap creation scenarios are to be disentangled. A combination
of spatially resolved submillimeter imaging, ultimately with the Atacama 
Large Millimeter Array (ALMA), and high resolution infrared and
millimeter-wave spectroscopy to trace the gas content will provide the 
key observations to drive our future understanding of these fascinating objects.

\bibliographystyle{apj}




\acknowledgments {The Submillimeter Array is a joint project between
the Smithsonian Astrophysical Observatory and the Acadenia Sinica
Institue of Astronomy and Astrophysics and is funded by the
Smithsonian Insitution and the Academia Sinica.}

\clearpage


\begin{deluxetable}{lccccccl}
\tablecolumns{8}
\rotate
\setlength{\tabcolsep}{0.08in}
\tablewidth{0pt}
\tabletypesize{\normalsize}
\tablecaption{\label{table:smaobs}Summary of SMA observations}
\tablehead{\colhead{Source} & \colhead{R.A.} &
\colhead{Dec.} & \colhead{Beam size} & \colhead{SSB T$_{\rm sys}$} & \colhead{Phase} & \colhead{Flux} & \colhead{Observation dates} \\
\colhead{} & \colhead{(J2000)} &
\colhead{(J2000)} & \colhead{} & \colhead{(K)} & \colhead{Calibrator} & \colhead{Calibrator} & \colhead{}}
\startdata
LkH$\alpha$ 330 & 03:45:48.28 & +32:24:11.8 & 0\farcs28x0\farcs33 &
211-870  & 3C111 & Titan & 2006 November 11\\
& & & &
210-685  & 3C111 & Uranus & 2006 November 19\\
SR 21N & 16:27:10.28 & -24:19:12.7 & 0\farcs42x0\farcs30 & 180-490 & 1625-254 & Callisto & 2007 June 10\\
HD 135344B & 15:15:48.43 & -37:09:16.22 & 0\farcs47x0\farcs25 & 170-610 & 1454-377 & Callisto & 2007 May 27 \\
& & & & 150-424 & 1454-377 & Callisto & 2007 June 8

\enddata
\end{deluxetable}

\begin{deluxetable}{lcccccccclc}
\tablecolumns{11}
\rotate
\setlength{\tabcolsep}{0.05in}
\tablewidth{0pt}
\tabletypesize{\footnotesize}
\tablecaption{\label{table:smaimages}Summary of the SMA imaging results}
\tablehead{\colhead{Source} & \colhead{F$_{\rm total}$} &
\colhead{RMS Noise} & \colhead{R$_{\rm hole}$}  & \colhead{R$_{\rm SED\,\, hole}$}  & \colhead{R$_{\rm disk}$$^1$} & \colhead{F$_{\rm
    hole}$} & \colhead{M$_{\rm hole}$} & \colhead{Inclina-} & \colhead{Null} & \colhead{Position} \\
\colhead{} & \colhead{(mJy)} &
\colhead{(mJy/beam)} & \colhead{(AU)} & \colhead{(AU)} & \colhead{(AU)} & \colhead{(mJy)}  & \colhead{(10$^{-4}$
  M$_\odot$)} & \colhead{tion ($^{\rm \circ}$)}& \colhead{(k$\lambda$)} & \colhead{angle ($^\circ$)}}
\startdata
LkH$\alpha$ 330 & 59.8  &  2.3 & 47  & 50 & 125 & 2.3 & 1.3$\pm$1.3 & 42 & 240-260 & 165\\
SR 21N & 217 & 3.7 & 33 & 18 & 140 & 16.7 & 7.5$\pm$0.9 & 21 & 200-240 & 15 \\
HD 135344B & 314 & 4.5 & 39 & 45 & 125 & 6.7 & 1.1$\pm$0.8 & 21 & 160-200 & 55 
\enddata
\tablenotetext{1}{R$_{\rm disk}$ is a lower limit to the outer disk radii measured where the disk emission is 3$\sigma$ above the noise. Due to the lower optical depths in the outer disk, this high resolution imaging is likely not sensitive to the outermost reaches of the disk.}
\end{deluxetable}




\end{document}